\documentclass[conference, 10pt]{IEEEtran}
\IEEEoverridecommandlockouts
\usepackage{cite}
\usepackage{amsmath,amssymb,amsfonts}
\usepackage{algorithmic}
\usepackage{graphicx}
\usepackage{svg}
\usepackage{textcomp}
\usepackage{xcolor}
\usepackage{authblk}
\usepackage{amsmath}
\usepackage[export]{adjustbox}

\usepackage{paralist}

\usepackage{graphicx}
\ifCLASSOPTIONcompsoc
    \usepackage[caption=false, font=normalsize, labelfont=sf, textfont=sf]{subfig}
\else
\usepackage[caption=false, font=footnotesize]{subfig}
\fi

\usepackage{todonotes}

\def\BibTeX{{\rm B\kern-.05em{\sc i\kern-.025em b}\kern-.08em
    T\kern-.1667em\lower.7ex\hbox{E}\kern-.125emX}}
    
\IEEEoverridecommandlockouts
\IEEEpubid{\makebox[\columnwidth]{978-1-7281-9829-3/20/\$31.00 \copyright~2020~IEEE }
\hspace{\columnsep}\makebox[\columnwidth]{ }}
    
\usepackage[bottom=1.0in,top=0.704in,right=0.68in,left=0.68in]{geometry}

\begin{document}

\title{When is Enough Enough? ``Just Enough" Decision Making with Recurrent Neural Networks for Radio Frequency Machine Learning\\
\thanks{The work of M. Moore was supported in part by the
Bradley Masters Fellowship through the Bradley Department of Electrical
and Computer Engineering at Virginia Tech.}
}

\author[1]{Megan Moore}
\author[2]{William H. Clark IV}
\author[3]{R. Michael Buehrer, PhD}
\author[4]{William C. Headley, PhD}
\affil[1,2,4]{Ted and Karyn Hume Center for National Security and Technology, Virginia Tech}
\affil[3]{Bradley Department of Electrical and Computer Engineering, Virginia Tech}

\maketitle

\begin{abstract}
 Prior work has demonstrated that recurrent neural network architectures show promising improvements over other machine learning architectures when processing temporally correlated inputs, such as wireless communication signals. Additionally, recurrent neural networks typically process data on a sequential basis, enabling the potential for near real-time results. In this work, we investigate the novel usage of "just enough" decision making metrics for making decisions during inference based on a variable number of input symbols. Since some signals are more complex than others, due to channel conditions, transmitter/receiver effects, etc., being able to dynamically utilize just enough of the received symbols to make a reliable decision allows for more efficient decision making in applications such as electronic warfare and dynamic spectrum sharing. To demonstrate the validity of this concept, four approaches to making "just enough" decisions are considered in this work and each are analyzed for their applicability to wireless communication machine learning applications.
 
\end{abstract}


\section{Introduction}{\label{sect:introduction}}
Recurrent Neural Networks (RNN) are a class of deep learning algorithms that particularly excel at processing inputs that have temporal correlation \cite{r12}. The primary driver for this capability is that RNNs rely heavily on utilizing memory structures in which the outputs of the network are based not only on the current input to the network but past inputs as well. This gives them the ability to learn time dependent effects more efficiently. Given this, RNNs have often been used in machine translation and natural language processing applications. Recently, RNNs have shown incredible promise for their usage in Radio Frequency Machine Learning (RFML) applications such as RF fingerprinting \cite{r7}, spectrum prediction with Cognitive Radios \cite{r8, r9}, and modulation classification \cite{r4, r5, r6}, among others.

Although the use of RNNs in these applications has been shown to improve performance over Convolutional Neural Network (CNN) architectures, there are tradeoffs. Since the inputs are processed sequentially rather than in a large batch, RNNs are typically very slow to train and evaluate \cite{r10}. To leverage the benefits of both approaches, and reduce their respective drawbacks, hybrid networks like WaveNet \cite{r1} and Fast WaveNet \cite{r2} have been developed that use dilation to mimic RNNs while reducing training and evaluation time. Other researchers have also focused on minimizing network complexity for use on resource constrained devices \cite{r11}.

In addition to training time, RNNs can also suffer from long evaluation times during inference. In many spectrum sensing applications, reliable decisions need to be made as quickly as possible. For example, in a cognitive electronic warfare system, the matter of a few milliseconds may be vital. This work introduces a concept by which to reduce the inference times of RNNs through dynamically ingesting ``just enough" input data to make reliable decisions.
\begin{figure*}[t]
    \centering
  \subfloat[Softmax Output of a 0 dB BPSK Signal \label{1a}]{%
       \includegraphics[trim={0 2.1cm 0 2.5cm},clip,height=7cm,width=0.5\linewidth]{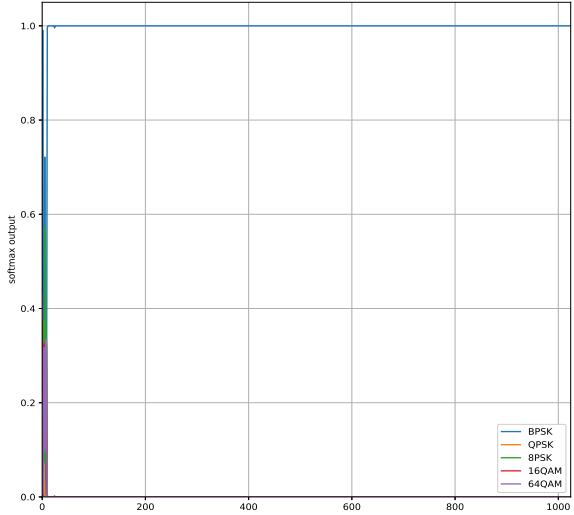}}
    \hfill
    \centering
  \subfloat[Softmax Output of a 0 dB 16-QAM Signal \label{1b}]{%
        \includegraphics[trim={0 2.1cm 0 2.5cm},clip,height=7cm,width=0.5\linewidth]{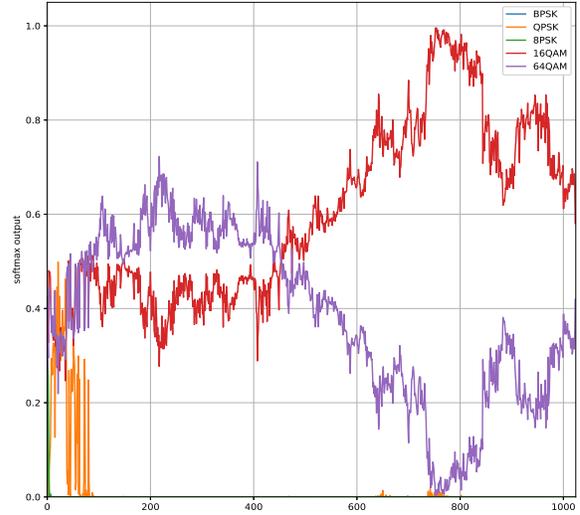}}
    \hfill

  \caption{The softmax output of two different 0 dB signals fed through Model 1 defined by Table \ref{model_param}. The output shown does not include any ``just enough" decision making. Instead, the decision made is the class with the highest output after the last received symbol is processed, here after 1024 input symbols. In (a) the network converges to its answer of BPSK after processing less than 10\% of the input symbols. In (b), the network struggles to converge to its answer of 16-QAM until around 80 percent of the input symbols have been processed. }
  \label{softmax} 
  \vspace{-1em}
\end{figure*}

\section{Motivation}\label{sect:motivation}
RFML-based RNNs are typically trained with signal inputs consisting of a pre-defined number of samples or symbols and the output is typically given after all inputs have been processed. To combat the slow training and evaluation times, smaller signal lengths can therefore be used. However, it is perhaps obvious that there is no one size fits all observation interval for the variety of signal and channel conditions that could be encountered. For example, longer observation intervals are required when differentiating between signal types such as 16-QAM and 64-QAM than for signal types such as BPSK and QPSK. This is true in both RFML-based and non-RFML-based spectrum sensing applications \cite{r13}. Figure \ref{softmax} shows an example of the output of a trained RNN (details of which will be discussed in Section III) for two signal types both at an SNR of 0dB. The x-axis represents the current input received symbol while the y-axis represents the RNN output after that symbol is processed. While the BPSK signal is recognized almost immediately, the 16-QAM signal is initially misidentified as 64-QAM and does not consistently produce the correct answer until much later (and even then with much less confidence).

In symbol-by-symbol implementations, RNNs can process more or less symbols than they are trained on. In the case of Figure \ref{softmax}, for example, the network was trained with a signal length of 1024 symbols, but clearly does not always need to process the entire trained signal length before making a confident output. In the BPSK case, the network identifies the correct signal type after processing less than 10 percent of the input symbols. In the 16-QAM case, it identifies the correct signal type after processing around 80 percent of the input symbols. 

Given the observed performance of the considered RNN architectures, the aim of this paper is to develop the concept of ``just enough" decision making that allows for utilizing signal observations less than the training size when processing simpler signals and potentially longer observations than used during training for more complex signals, the later of which is left as investigation in future work. More specifically, this paper investigates approaches by which the network can determine dynamically at what point it has enough confidence in its decision to stop processing new inputs. 

The scope of this paper's application is for an RFML-based Automatic Modulation Classification (AMC), but can be generalized to other applications of interest, such as those discussed in Section \ref{sect:introduction}. In Section \ref{sect:systemmodel}, the considered AMC and its representative RNN architectures are presented. In Section \ref{sect:techniques}, the ``just enough" decision techniques investigated are introduced and their comparative performance is presented in Section \ref{sect:analysis}. Finally, this work is concluded and future work is suggested in Section \ref{sect:conclusions}.

\section{System Model}{\label{sect:systemmodel}}
In order to better generalize the takeaways of the concepts introduced within this work, a total of five different RNN based AMCs were trained and evaluated. Since both Long-Short Term Memory (LSTM) and Gated Recurrent Unit (GRU) based RNNs have seen widespread use in RFML applications, both architectures are considered. 

\subsection{RNN Training and Evaluation Data}
 To facilitate model training, synthetic (i.e. I/Q) received symbols were generated for five modulation schemes, namely BPSK, QPSK, 8-PSK, 16-QAM, and 64-QAM. A fixed signal length of 1024 symbols was used for all generated data. Symbols were pulse shaped with a root-raised-cosine filter using a roll-off factor of 0.35 and an over-sampling rate of 4 samples per symbol. The pulse-shaped samples were then fed through an AWGN channel with a variable integer SNR from 0 to 10 dB. To produce the received symbols used as input to the RNNs under test, time/frequency synchronization and a matched filter is assumed. Each RNN considered was trained on 8,000 examples per SNR per class, for a total of 88,000 per class, and tested on a separately generated dataset of 100 examples per SNR per class, for a total of 1,100 per class. 

\subsection{Model Architectures}
The basic architecture for each model is shown in Tables \ref{model_arch} and \ref{model_param} and were trained and evaluated using the PyTorch library. As can be seen, to better generalize the takeaways of this work across RNN architectures, different values were chosen for the number of recurrent layers, the hidden size of the layers, and the size of the linear layers. Dropout regularization of 0.5 was used for all models. Finally, Table \ref{model_param} also shows the average probability of correct classification (PCC) of each model evaluated on the testing set for the full input size of 1024 symbols. \emph{Note: Figure \ref{softmax} discussed within Section \ref{sect:motivation} was generated from Model 1}. 
\begin{table} [h]
\caption{Overall Architecture for the RNN Models Considered.}
\centering
{\small
\begin{tabular}{c  c  c  c} 
\textbf{Name} & \textbf{Layer} & \textbf{Activation} &  \textbf{Description}\\ \hline  \hline
Input & - &  - & input size of 2  \\ \hline 
Recurrent & variable & - & variable hidden size\\ \hline
Dropout & - & - & p=0.5  \\ \hline
Linear & 1 & ReLU & variable linear size \\ \hline
Output & 1 & Softmax & output size of 5 \\
\end{tabular}
}
\label{model_arch}
\end{table}

\begin{table} [h]
\caption{Architecture Parameters for each considered RNN Model.}
\centering
{\small
\begin{tabular}{c  c  c  c  c  c } 
\textbf{Model} & \textbf{Type} & \textbf{Layers} & \textbf{Hidden} & \textbf{Linear} & \textbf{PCC}\\ \hline  \hline
0 & LSTM & 2 & 128 & 64 & 0.86\\ \hline 
1 & LSTM & 3 & 64 & 32 & 0.87 \\ \hline
2 & GRU & 2 & 128 & 64 & 0.87 \\ \hline
3 & GRU & 3 & 64 & 32 & 0.87 \\ \hline
4 & GRU & 2 & 128 & 128 & 0.83 \\
\end{tabular}
}
\label{model_param}
\end{table}

\section{``Just Enough" Decision Techniques}{\label{sect:techniques}}
The concept of ``just enough" decision making is based on the idea that an RNN can be made more efficient by stopping its processing of new input once it is confident in its decision. Figure \ref{softmax}(a) showed that for a more simplistic input (here, BPSK) the RNN reached a decision almost immediately while Figure \ref{softmax}(b) showed that for a more complex input (here, 16-QAM) the RNN initially made a wrong prediction. While it is clear from this example, and others not shown here, that ``just enough" decision making is feasible by visual inspection, techniques by which to algorithmically determine these stopping conditions are necessary for real-world application. In this work, four candidate techniques are investigated in order to determine these stopping conditions, but should not be considered an exhaustive solution space for this proposed concept. Each candidate technique is a post-training technique that utilizes the softmax value of the RNN. No changes were necessary to the training or architecture of the RNNs. 

There are two important considerations that were made when investigating possible techniques. First, it is important that the considered technique can be implemented in parallel with the network and doesn't require changes to the model's architecture or training process. Secondly, and perhaps most importantly, the chosen techniques must be able to be calculated within the symbol evaluation time of the RNN. If a technique takes longer than this to execute, than it can't keep up with the RNNs execution, thus rendering the technique impractical for ``just enough" processing.

\subsection{Subset Technique} 
The first technique, and most simplistic, is referred to here as the subset (SUB) technique. As the name implies, this technique processes a user-defined subset of the original signal. More specifically, the amount of input processed before making its decision is determined by the user-defined variable \textit{duration}. For example, for a training signal length of 1024 symbols and a \textit{duration} of 100, the network will always process only the first 9.7 percent of the signal and stop. While this approach is not intended to be a realistic candidate technique, it acts as a simplistic baseline for comparative purposes.

\subsection{Threshold Technique}
The second technique, which will be referred to as the threshold (THR) technique, processes the input until the softmax value of the output exceeds a user-defined value termed simply \textit{threshold}. For example, if the value of \textit{threshold} is 0.9, then the classifier will stop as soon as the softmax output for any given class exceeds 0.9. 

\subsection{Subset-Above-Threshold Technique}
The third technique will be referred to as the subset above threshold (SAT) technique and is a combination of the previous two techniques. More specifically, this technique requires that the softmax value of a class exceed a user-defined value, termed \textit{threshold}, for a certain number of consecutive inputs, termed \textit{duration}. For example, if the value of \textit{threshold} is 0.9 and \textit{duration} is 100, then the softmax value for the highest class must exceed 0.9 for 100 consecutive symbols before the RNN will determine the output. 

\subsection{Delta Threshold Technique}
The fourth technique, which will be referred to as delta-threshold (DEL), is designed to look for stability or zero slope in the softmax value. While the SAT technique looks for the softmax output to remain above a certain value, DEL looks for the change in the output to stay within a certain range. This addresses cases where the network may not have a high confidence in its decision, but where new information through further processing will not change the network's decision. It depends on two user-defined variables: \textit{delta-threshold} which determines the amount of change between outputs that is tolerated and \textit{duration}, the number of consecutive inputs for which the change must be below the \textit{delta-threshold}. Two change values are used: the first is the change between the current input and the first input in the series and the second is the change between consecutive inputs. For example, if \textit{duration} is 100, the \textit{delta-threshold} is 0.1, and the first symbol in the series is 0.9, then the output of the RNN must stay between 0.8 and 1 for 100 symbols. In addition, the change between two consecutive symbols must not exceed 0.1, if the output drops from 0.97 to 0.8 the counter will reset. The network will only make its decision if both these conditions are true for the highest class for 100 consecutive symbols. 

\section{Analysis}{\label{sect:analysis}}
To quantify the efficacy of each candidate technique, two metrics are used to compare the performance and processing speed of the RNNs. The first metric is the RNN's average PCC, which is simply the number of times the correct signal class is predicted divided by the total number of signals evaluated. The second metric is the portion of the total input signal processed (PSP) before the RNN made its decision. 

Each technique described in Section \ref{sect:techniques} is evaluated on each model presented in Section \ref{sect:systemmodel} and then compared to the model's baseline PCC, when it does not leverage any of the candidate techniques. To determine the tradeoff between PCC and PSP, a range of values for \textit{threshold}, \textit{delta-threshold}, and \textit{duration} are evaluated. More specifically, \textit{thresholds} above 0.5 and \textit{delta-thresholds} below 0.5 and \textit{durations} from 100 to 900 are evaluated. For comparison purposes, values of \textit{duration} will be displayed as a percentage of the total training signal duration. 

\begin{figure*}[h!]
    \centering
  \subfloat[Subset Technique \label{1a}]{%
       \includegraphics[trim={0 1.25cm 0 2.25cm},clip, width=0.45\linewidth]{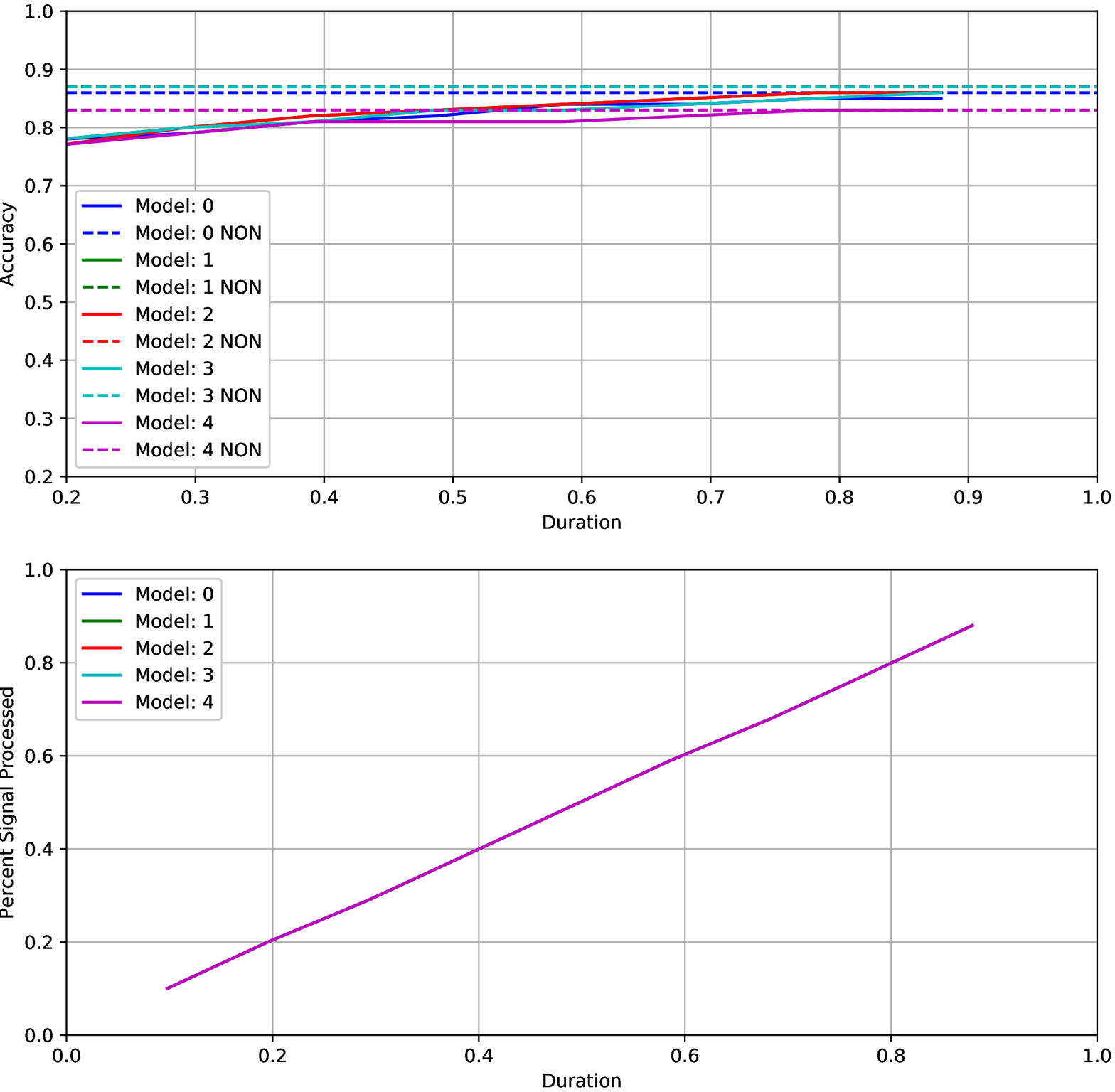}}
  \subfloat[Threshold Technique \label{1b}]{%
        \includegraphics[trim={0 1.25cm 0 2.25cm},clip,width=0.45\linewidth]{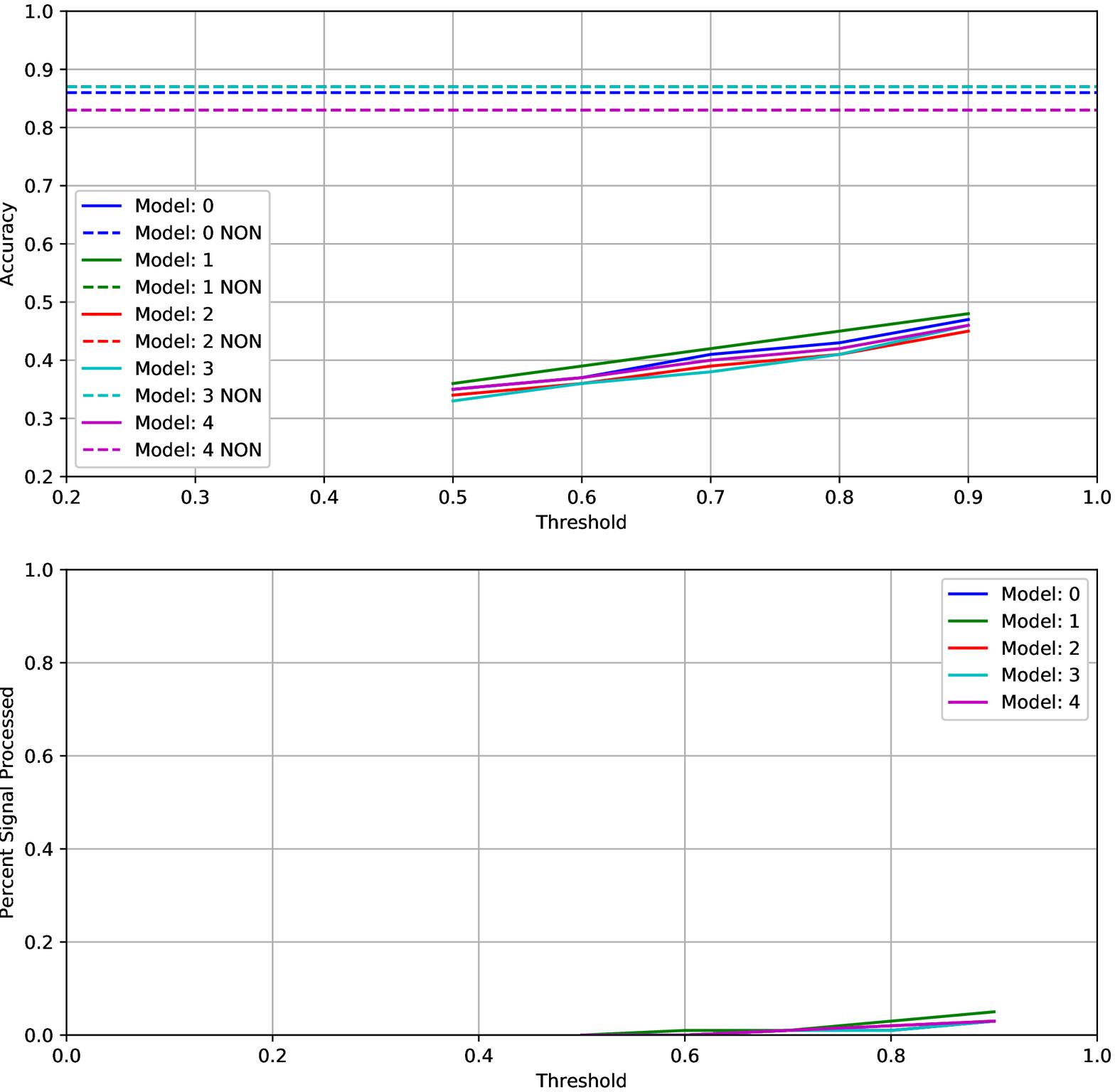}}
  \\[-2ex]
  \subfloat[Subset Above Threshold Technique \label{1c}]{%
        \includegraphics[trim={0 1.25cm 0 2.25cm},clip, width=0.45\linewidth]{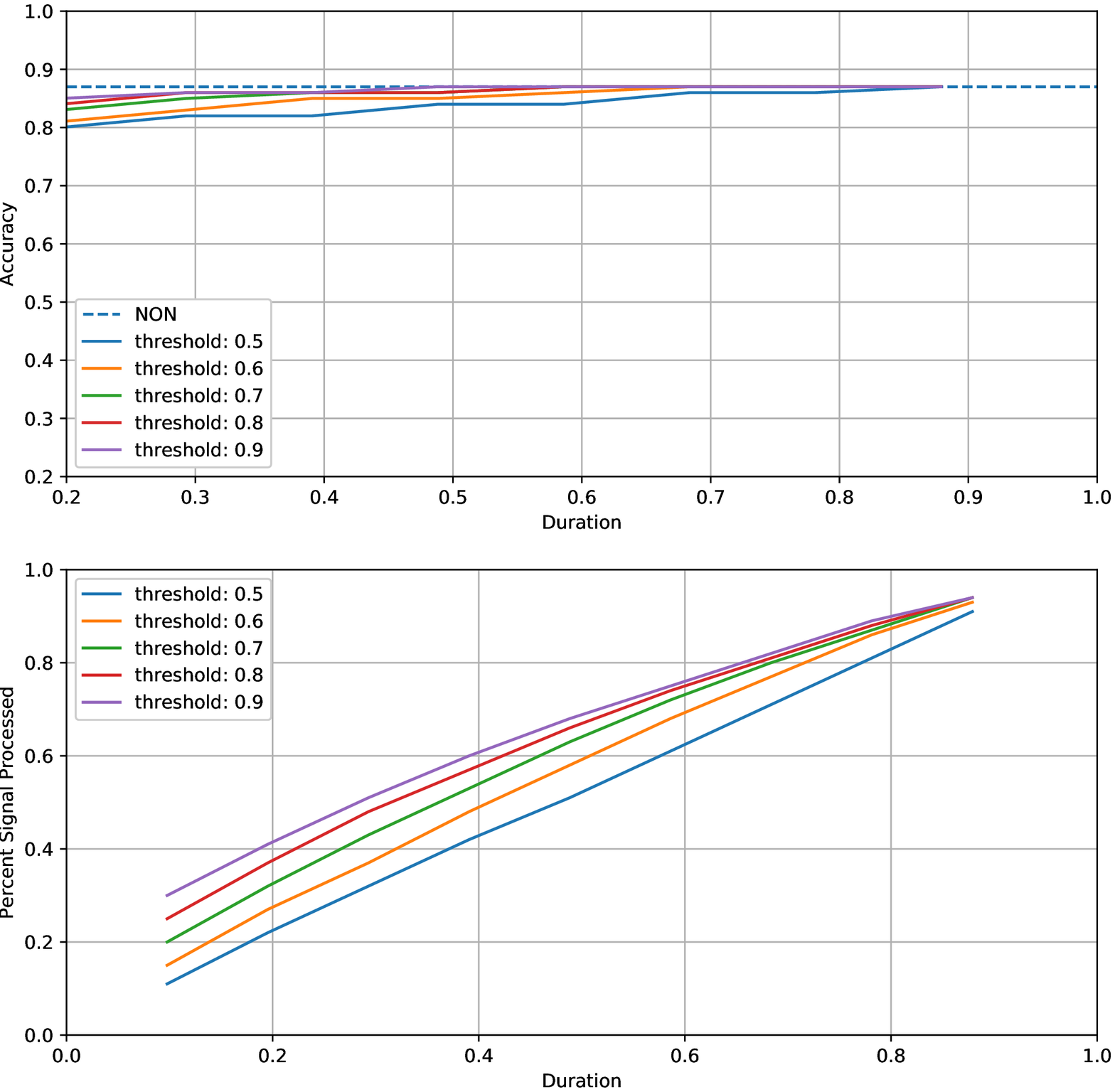}
        }
  \subfloat[Delta-Threshold Technique\label{1d}]{%
        \includegraphics[trim={0 1.25cm 0 2.25cm},clip,width=0.45\linewidth]{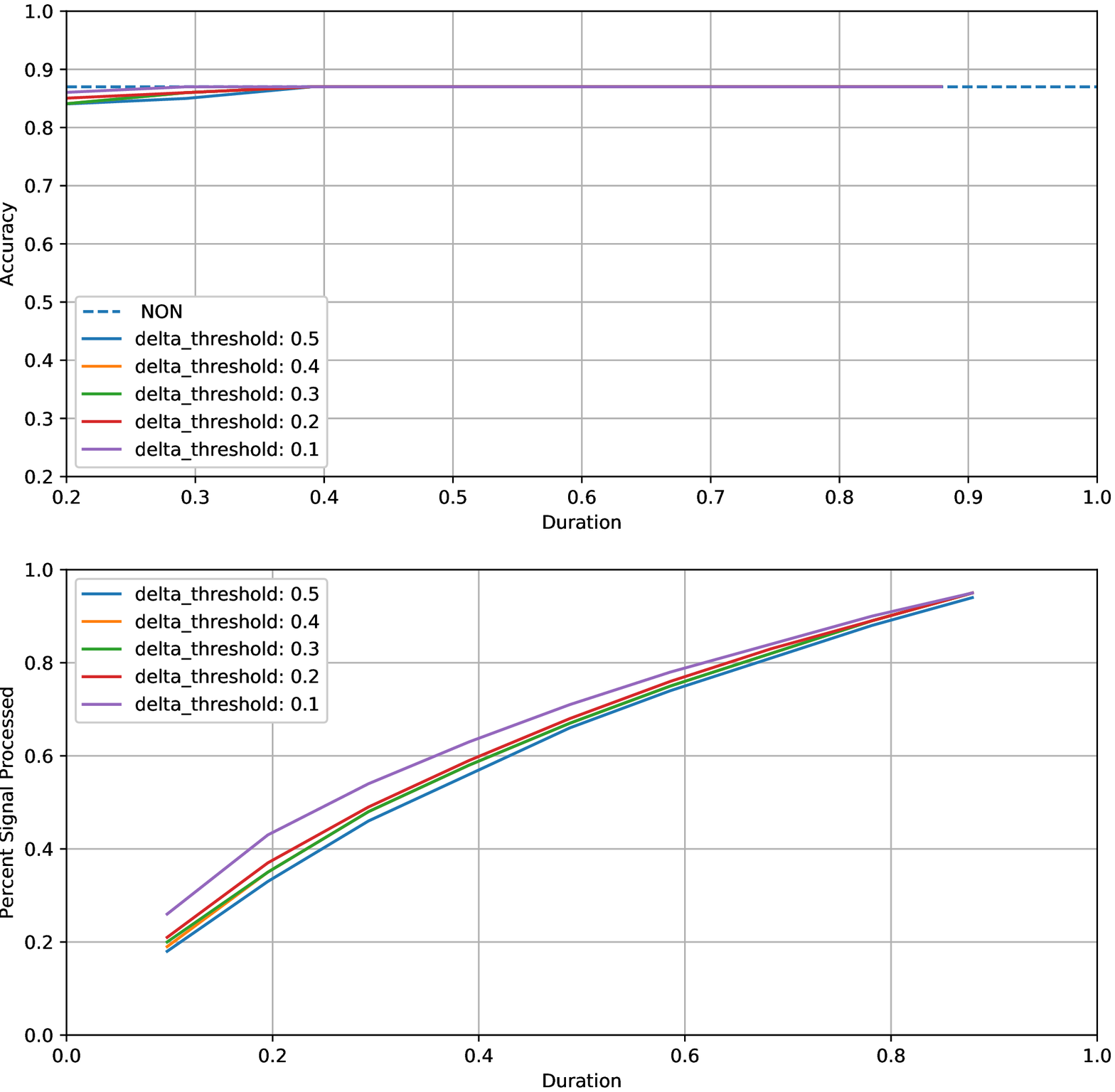}
        }
  \caption{A tradeoff analysis of each candidate technique was performed by testing multiple combinations of the user-defined variables. The top plot of each figure shows the performance of the network for that combination with the RNN model's baseline performance (at a PSP of 1) shown by the dotted line labeled NON. The bottom plots show the percent of the input processed for each combination with a maximum value of 1 meaning that the full input was processed. In (a) and (b), the results for each model are shown. In (c) and (d) results are only shown for Model 1 (with other models showing similar trends).}
  \label{tradeoff} 
  \vspace{-10pt}
\end{figure*}

Figure \ref{tradeoff} presents the chosen evaluation metrics for each considered ``just enough" decision technique. More specifically, Figures \ref{tradeoff}(a) and \ref{tradeoff}(b) respectively present the SUB and THR techniques evaluated for all considered RNN models defined by Tables \ref{model_arch} and \ref{model_param}, while Figures \ref{tradeoff}(c) and \ref{tradeoff}(d) respectively present the SAT and DEL techniques only for Model 1 (results for the other models showed similar trends and are omitted for clarity). In each subfigure, NON represents the performance of the stated RNN model without the use of any ``just enough" decision technique (in other words, the performance of the stated model at a PSP of 1). 

Table \ref{best_performance} presents the best performance for each considered technique applied to each considered model by determining the lowest PSP with an average PCC equal to the original networks performance. For the DEL technique, the Threshold column represents the variable \textit{delta-threshold} rather than \textit{threshold}. While Table \ref{best_performance} only displays results that had an average PCC equivalent to the original baseline, if some performance loss is acceptable than the percent of the signal processed can be reduced even further. By accepting a performance loss of 1 percent, Model 3 processed 16 percent less compared to the results shown in Table \ref{best_performance} for the SAT method and 25 percent less for the DEL method. 

In the following, the results presented in Figure \ref{tradeoff} and Table \ref{best_performance} are discussed for each of the considered ``just enough" decision techniques.

\subsection{Subset}
The SUB technique gives the most control over the number of input processed, as shown in the linear curve of Figure \ref{tradeoff}(a). The major drawback of this technique is that it doesn't leverage any information on the output of the network, therefore, it is entirely inflexible. This results in a significant performance loss for lower values of \textit{duration}. With a \textit{duration} of 200, Model 0 saw accuracy comparable to the original model when processing BPSK signals. However, for 64-QAM signals, performance decreased by 11 percent and for 16-QAM, it decreased by 17 percent. These results mirror the example illustrated in Figure \ref{softmax} and show the need for  flexibility. Although the SUB technique is not analogous to training on a smaller signal length, the results shown suggest that processing the same number of symbols for every signal type is not the best path forward. Future work will include training on a smaller signal length and then evaluating for a larger duration using some of the techniques defined in this paper.

\subsection{Threshold}
The THR technique overall performed very poorly. While it processed significantly less of the input symbols (less than 10 percent), the accuracy for each model was considerably lower than its original baseline. A deeper understanding of the softmax layer and classification networks is useful to understanding why this is the case. The softmax layer forces the values of each output to sum to one, similar to a discrete probability distribution. Due to this, the softmax value is often, perhaps incorrectly, used as a measure of the network's confidence in its decision. By this logic, when an RNN processes the first input symbol of a signal each softmax value should be approximately equal (around 0.2). In practice however, one class will typically have a significantly higher softmax value than the others, even for the first few symbols when the RNNs memory states are still being initialized. 

If softmax was a true measure of the network's confidence, then the THR technique would likely work very well. However, instead the technique causes the network to make a decision before it has enough information. Figure \ref{tradeoff}(b) shows that while the average PCC of the THR technique was low, it did increase as the value of the \textit{threshold} increased, to a point. For example, for Model 4, with a \textit{threshold} of 0.99, the network processed less than 10 percent of the signal and had an average PCC of only 0.54. 

\subsection{Subset Above Threshold}
As a combination of the two previous techniques, the SAT technique performed very well. It avoided the primary issues of the prior two techniques by processing at least a set number of input and by incorporating information on the output of the RNN. This technique was able to achieve the baseline accuracy of the original models while processing an average of 20 to 40 percent less of the input symbols. The overall best combination for this technique for each model is shown in Table \ref{best_performance}. The combinations resulting in the best performance while minimizing the percent of the input symbols processed tended to use a \textit{duration} of around 0.5 and a higher \textit{threshold} such as 0.8 or 0.9. Other combinations with lower values of \textit{threshold} and higher values of \textit{duration} achieved comparable performance, but processed a larger percentage of the signal. 

\subsection{Delta Threshold }
Similar to the SAT technique, the DEL technique avoids the pitfalls of the SUB and THR techniques. However, unlike the SAT technique, the DEL technique was designed to focus on stability. This technique was able to achieve the baseline accuracy of the original models while processing an average of 35 to 45 percent less of the input symbols. Figure \ref{tradeoff}(d) shows the expected trends with a smaller \textit{delta-threshold} resulting in better performance and a larger percentage of the input processed. However, there is not a significant difference between performance for the different \textit{delta-threshold} values while there is a large gap in the PSP at lower values of \textit{duration}. To achieve the performance of the original model, each \textit{delta-threshold} needed a \textit{duration} of around 0.4. The DEL technique had the best overall performance as well as the least dependency on the user-defined variables. While some initial choices in variables for the other techniques resulted in very poor performances, the DEL technique rarely had an average PCC of less than 80 percent. 

\begin{table}
\caption{Best Performance for each considered RNN Model.}
\centering
{\small
\begin{tabular}{c  c  c  c  c  c } 
\textbf{Model} & \textbf{Technique} & \textbf{PCC} & \textbf{PSP}& \textbf{Duration}& \textbf{Threshold}\\ \hline  \hline
0 & SUB & 0.86 & 0.88 & 0.88 & -\\ \hline 
0 & SAT & 0.86 & 0.58 & 0.39 & 0.9\\ \hline
0 & DEL & 0.86 & 0.57 & 0.39 & 0.4\\ \hline
1 & SUB & 0.86 & 0.78 & 0.78 & -\\ \hline 
1 & SAT & 0.87 & 0.68 & 0.49 & 0.9\\ \hline
1 & DEL & 0.87 & 0.54 & 0.29 & 0.1\\ \hline
2 & SUB & 0.83 & 0.78 & 0.78 & -\\ \hline 
2 & SAT & 0.87 & 0.67 & 0.49 & 0.9\\ \hline
2 & DEL & 0.87 & 0.63 & 0.39 & 0.2\\ \hline
3 & SUB & 0.86 & 0.88 & 0.78 & -\\ \hline 
3 & SAT & 0.87 & 0.75 & 0.59 & 0.9\\ \hline
3 & DEL & 0.87 & 0.76 & 0.59 & 0.1\\ \hline
4 & SUB & 0.83 & 0.78 & 0.78 & -\\ \hline 
4 & SAT & 0.83 & 0.66 & 0.59 & 0.6\\ \hline
4 & DEL & 0.83 & 0.50 & 0.29 & 0.1\\ \hline
\end{tabular}
}
\label{best_performance}
\end{table}

\begin{figure*}[h!]
    \centering
  \subfloat[Analysis by Signal Type\label{1a}]{%
       \includegraphics[trim={0 1.6cm 0 2.25cm},clip,width=0.5\linewidth]{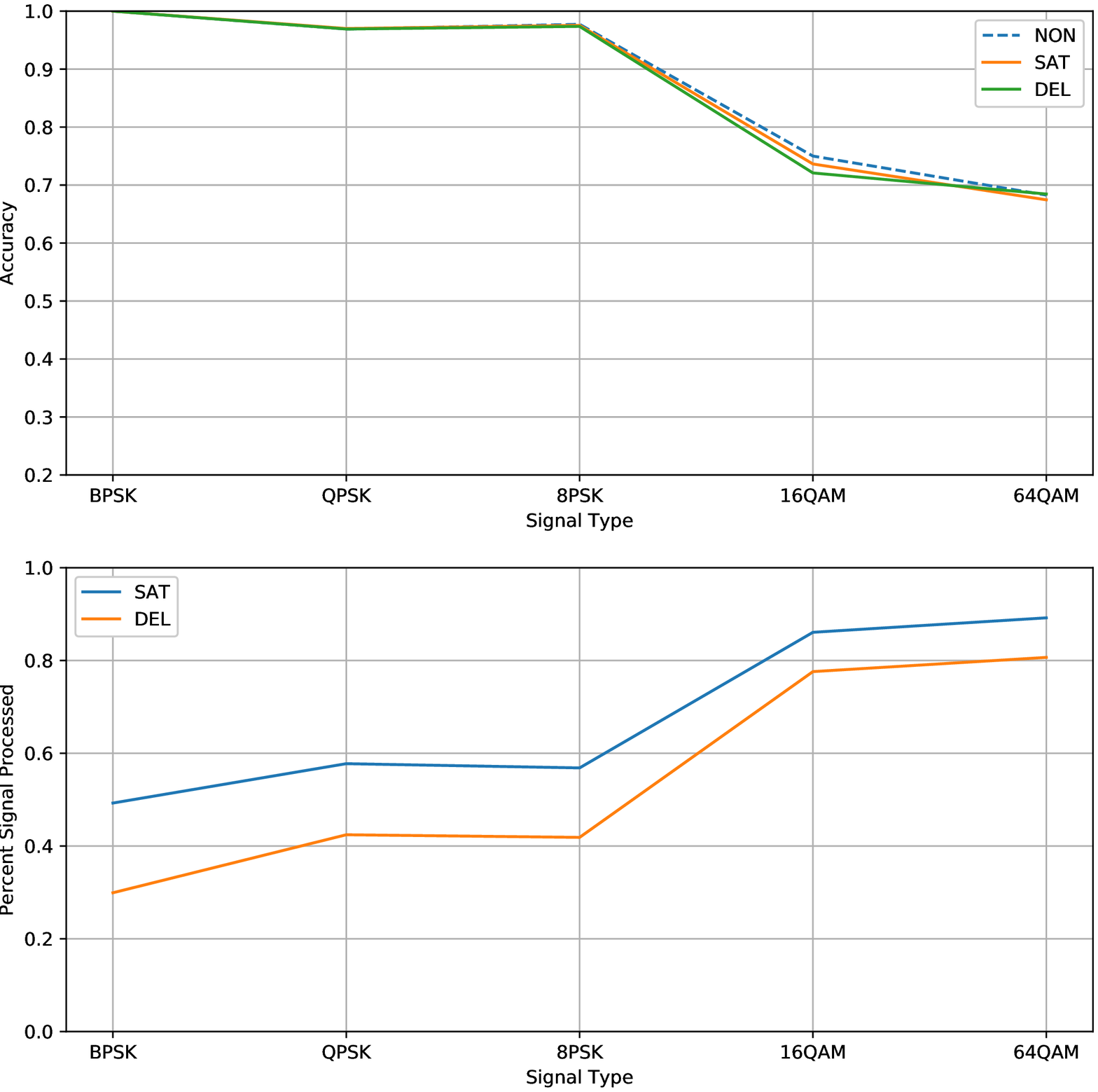}}
  \subfloat[Analysis by SNR\label{1b}]{%
        \includegraphics[trim={0 1.6cm 0 2.25cm},clip,width=0.5\linewidth]{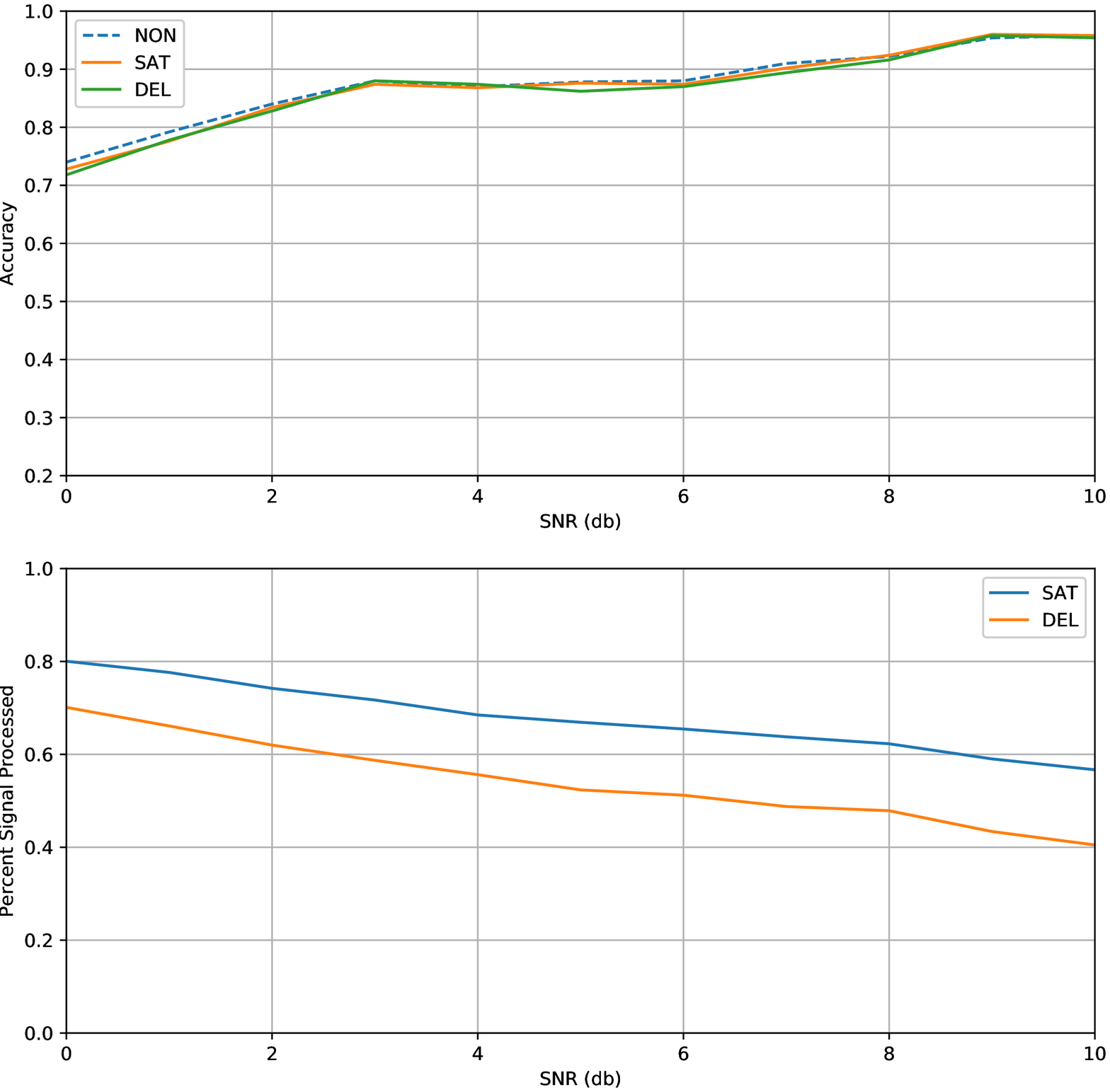}}
    \hfill
  \caption{A breakdown analysis of RNN Model 1 was performed by testing the SAT and DEL techniques on different input cases. The \textit{duration}, \textit{threshold}, and \textit{delta-threshold} were chosen based on Table \ref{best_performance}. The top plots show the accuracy of the network with "just enough" decision making while the dotted NON line represents the network performance after processing the full input duration, (i.e. at a PSP of 1). The bottom plots show the percentage of the input symbols processed in each case. In (a) each signal type was tested individually to see if the``just enough" techniques processed more of the input signal for higher order modulation schemes. In (b) each integer SNR value was tested to see if the "just enough" techniques processed more of the input signal for lower values of SNR.}
  \label{breakdown} 
\end{figure*}

\subsection{Investigation of Flexibility}
To check the flexibility of the different techniques, a breakdown analysis was performed to check whether the network was making a decision earlier for simpler signal types and higher SNRs, as would be expected. Figure \ref{breakdown}(a) shows the average PCC and PSP of Model 1 while using the SAT and DEL techniques for each signal type and Figure \ref{breakdown}(b) shows the same results for each SNR. The \textit{duration}, \textit{threshold}, and \textit{delta-threshold} were chosen based on Table \ref{best_performance}. The RNN processed 40 percent less of the BPSK signals than the 64-QAM signals when using the SAT technique. When using the DEL technique, the network processed 50 percent less. The variation due to SNR was less pronounced, but still exhibited a clear trend. The SAT technique processed 23 percent less of the input symbols at an SNR of 10 dB than at an SNR of 0 dB and the DEL technique processed 30 percent less. Clearly, both the SAT and DEL techniques are processing a larger portion of the input symbols for lower SNRs and higher order modulation schemes. 

\section{Conclusion and Future Work}{\label{sect:conclusions}}

Although RNNs are known to perform well at tasks in RFML, they struggle in symbol-time implementation due to their need to process each symbol sequentially. However, due to this, the networks are able to evaluate a variable number of symbols. Four different techniques were used to determine when the network should stop processing new input symbols and display an output. Of particular interest was whether the techniques could implicitly learn to differentiate between simplistic signals (higher SNRs and simple modulation formats such as BPSK) and more complex ones. The SUB technique processed a user-defined number of symbols and therefore was not able to differentiate between signal cases. However, the SAT and DEL techniques were shown to process significantly less of the signal for lower order modulation schemes and higher SNRs. This paper has demonstrated that ``just enough" decision making techniques allow an RNN to process less of the input for more simplistic signals without affecting the RNN's performance. The flexibility gained with this approach allows simpler signal types to be identified quickly while more complex signal types can be processed for a longer period. 

All networks in this paper were trained on received symbols, however, this requires strong assumptions on time/frequency synchronization as well as pre-knowledge of the pulse-shaping filter. Future work will include comparing networks trained with samples to those trained with symbols. The signal length of 1024 allowed for evaluation for smaller input lengths. However, it is reasonably common to find networks trained and evaluated on a smaller signal length, like 128. While the results from the SUB method suggest that evaluating all input signals on the same, smaller length will degrade performance for more complex signal types, the same may not be true for networks designed to operate for a smaller signal length. As such, future work will include training on smaller signal lengths like 128, and evaluating them for larger input length. 

While implementing a ``just enough" decision making technique, like those considered within this work, in symbol-time is beyond the scope of this paper, an implementation could be valuable for multiple reasons. When the duration and complexity of the signals of interest vary, the network can provide faster results for simpler cases. This would be particularly beneficial in time sensitive applications like electronic warfare since the network will process only as much information as it needs to make its decision. The technique may also offer an additional method of instilling user-confidence in a decision since the output must meet user-defined criteria before producing a decision. Additionally, the ``just enough" decision making techniques described in this work were used exclusively in the evaluation stage. However future work could develop a more complex method of incorporating the ``just enough" decision making into an RNN's training process. 

\bibliographystyle{IEEEtran}
\IEEEtriggeratref{5}
\bibliography{man}
\end{document}